\documentclass[acmtog, authorversion, nonacm]{acmart} %

\AtBeginDocument{%
  \providecommand\BibTeX{{%
    \normalfont B\kern-0.5em{\scshape i\kern-0.25em b}\kern-0.8em\TeX}}}

\usepackage{subcaption}
\usepackage{wrapfig}

\setcopyright{acmcopyright}
\copyrightyear{2024}
\acmYear{2024}
\acmDOI{XXXXXXX.XXXXXXX}

\begin{document}

\title[Imagining a Future of Designing with AI]{Imagining a Future of Designing with AI: Dynamic Grounding, Constructive Negotiation, and Sustainable Motivation}

\author{Priyan Vaithilingam}
\authornote{Equal contributions.}
\email{pvaithilingam@g.harvard.edu}
\affiliation{%
  \institution{Harvard University}
  \city{Cambridge}
  \state{Massachusetts}
  \country{USA}
  \postcode{02134}}

\author{Ian Arawjo}
\authornotemark[1]
\email{ian.arawjo@umontreal.ca}
\affiliation{%
  \institution{Universit\'{e} de Montr\'{e}al}
  \city{Montr\'{e}al}
  \state{Quebec}
  \country{Canada}
  \postcode{H3T 1J4}}

\author{Elena L. Glassman}
\email{glassman@seas.harvard.edu}
\affiliation{%
  \institution{Harvard University}
  \city{Cambridge}
  \state{Massachusetts}
  \country{USA}
  \postcode{02134}}

\renewcommand{\shortauthors}{Vaithilingam, Arawjo, \& Glassman}

\newcommand{\sayit}[1]{``\textit{#1}''}
\newcommand{\say}[1]{``#1''}

\begin{abstract}
  We ideate a future design workflow that involves AI technology. Drawing from activity and communication theory, we attempt to isolate the new value large AI models can provide design compared to past technologies. We arrive at three affordances---\emph{dynamic grounding}, \emph{constructive negotiation}, and \emph{sustainable motivation}---that summarize latent qualities of natural language-enabled foundation models that, if explicitly designed for, can support the process of design.
  Through design fiction, we then imagine a future interface as a diegetic prototype, the story of \emph{Squirrel Game}, that demonstrates each of our three affordances in a realistic usage scenario. Our design process, terminology, and diagrams aim to contribute to future discussions about the relative affordances of AI technology with regard to collaborating with human designers. %
\end{abstract}


\maketitle

\section{Introduction}

The advent of AI deep learning techniques and foundation models (commonly referred to as large language models or LLMs) marks a paradigm shift in human-computer interaction (HCI) research \cite{bommasani2021opportunities}. Chatting with foundation models like GPT-4-vision, for instance, people can now submit whiteboard drawings to generate code for open-domain tasks, interactions that previously required custom models and substantial feature engineering \cite{hammond2007ladder}.  
Yet it remains unclear what \emph{unique} value natural-language-enabled foundation models can bring to design processes compared to past technologies---and, more specifically, to designing new tools to support design. %

In this paper, we imagine how AI systems might support a user through the process of design. We establish concepts, terminology, and diagrams that can help ground further conversations in the community when discussing human-AI collaborative systems that support design. %
To begin, we define three unique affordances of LLMs compared to past technologies: \emph{dynamic grounding,} \emph{constructive negotiation,} and \emph{sustainable motivation}. Each relates to how humans successfully communicate and sustain participation over joint projects. These affordances are aspirational, and far from being used well or at all in current AI interfaces. %
We also frame the promises of AI as \emph{shifting power dynamics} \cite{li2023beyond} between machines and humans with respect to grounding communication, and by reducing \emph{translation work} \cite{arawjo2020write}, or the degree to which humans must ``submit'' to a software's interactional and representational expectations.

Through design fiction and scenario-based design \cite{bleecker2022design, nathan2008envisioning}, we then realize these affordances in a diegetic prototype~\cite{diegetic-prototype}---the story of \emph{Squirrel Game}, where a child interacts with a fictional AI game design tool called \emph{Game Jammer} through a pen-based tablet, collaboratively creating a 2D game with a squirrel protagonist (Section~\ref{story}). Scenario-based design is a method where one creates fictional albeit realistic stories centered around imagined technologies for the purpose of spurring further design and discussion \cite{linehan2014alternate, rosson2012scenario}. %
The story of Squirrel Game depicts an extended interaction in order to showcase how each of our three affordances appears in a real-world scenario.

We arrived at this work through weeks of sketching, imagining, and prompt-prototyping together on a whiteboard, over a page or screen. %
Our goal was to refrain from our tendency as HCI systems researchers to jump into implementation, which though exciting can narrow our attention, inhibit our imaginations, and incentivize us towards familiar designs \cite{willis2006ontological, arawjo2020write, bleecker2022design, rosson2012scenario}. %
We wanted to slow down and think carefully before we took the next step in human-AI interaction. Thus, our contribution is chiefly conceptual. In Section~\ref{story-debrief} we overview some technical implications and terminology necessary to make Game Jammer a reality. %

\section{Three Affordances of AI for Design}

We began our investigations by asking: \emph{What new value can natural language AI models provide to design processes that was difficult or impossible to achieve with classical methods?} In our discussions, we drew from activity theory, a framework of human activity and development prominent in early HCI research \cite{kaptelinin2006acting}, as well as \emph{Using Language}, a seminal work by Clark on how humans collaborate over joint projects \cite{clark1996using}. We isolated AI's potential for dynamic grounding, constructive negotiation, and sustainable motivation as three key affordances. All three relate to the fact that foundation models (a) embed a wide degree of cultural and social context outside of a specific domain, (b) can apply this understanding to augment outputs within a context, (c) can engage in natural dialogue with users that adapts to their notation and linguistic preferences. %

Here, we motivate and define these affordances, citing relevant literature. Each affordance may serve as a fruitful resource when asking how AI can support design processes; however, they may not be utilized well or at all by current interfaces. They are also not the \emph{only} affordances that can be useful in the context of human-AI collaboration.

\subsection{Dynamic Grounding} 

A long line of work in linguistics, following Grice and Clark \cite{clark1996using, grice1975logic}, views communication as a cooperative game where both the entities (the speaker and the listener) try to understand each other to succeed. For the two entities to successfully communicate, they need to coordinate not only in the semantics but also share a common representation or lingo---what we may call a \emph{notation.} Notations ground communication, both between people and between people and machines. People ground their interpersonal communication in ad-hoc ways, e.g., using commonly agreed-upon terminologies, sketching a diagram, jotting a note, prototyping an interface. By defining and referencing visual, linguistic, and interactive notations, people mutually establish \emph{common ground} when working together on joint projects \cite{clark1996using}. 

Take programming for example. In accomplishing a goal with a programming language (PL), the user is both the designer and the designed: at first, the power dynamic is entirely on the PL's side---the user must adopt the notation of the tool. Thus, initially the programming language "grounds" the user. But, over time the user builds their own notation into the tool by creating functions, APIs, libraries, etc. which reduces the distance between their \textit{preferred} means of communicating and the fixed notation of the tool---the user grounds the machine. %

Although humans \emph{can} ground communication with the machine in their preferred way, then, it first requires learning the interface's language, a process demanding tremendous time, knowledge, and effort. In human-computer communication, common ground has therefore traditionally been dictated by the software or hardware (and indirectly, by its designers) and is mostly fixed. Said differently, for a user to communicate with the computer successfully, they are required to submit to the machine's power and the power of its designers \cite{li2023beyond}. This requirement to learn the notation has colloquially been dubbed  the \textit{learning curve}~\cite{soloway1994learner} and has been extensively studied by the HCI community. Over the past few decades, we have come a long way in flattening the learning curve by making shared notations and interactions intuitive and easy to learn for most users. However, it is neither possible to remove the learning curve nor possible to optimize the tool for every user who may come from vastly different socio-cultural backgrounds. Hence, notations and interactions are still largely centered around the anticipated average user, and the onus of learning lies on the user.

Foundation models promise to reverse the power dynamic. %
Large AI models are trained on vast amounts of human data can make them capable of interpreting a user's ad-hoc notations. %
Tools like DynaVis~\cite{vaithilingam2024dynavis} enable users to dynamically generate hyper-contextual, ad-hoc user interfaces to communicate with the tool. Academics have introduced terms to refer to this quality of large models, such as Litt's ``malleable software''~\cite{litt2023malleable}, ``bespoke interfaces,'' by \citet{vaithilingam2019bespoke}, or ``dynamic interfaces'' used by Google AI in a recent address~\cite{google2023gemini}. 
However, these terms do not describe the \emph{affordance} of the design material itself, but rather \emph{outcomes} from its existence. To describe the affordance of a thing requires articulating what value it provides humans in the world~\cite{vyas2006affordance}.\footnote{The term affordance has long been contested and has myriad definitions \cite{osiurak2017affordance}. However, shared across definitions is the notion that an affordance arises from human \emph{relationship} to the artifact during human activity. We align with Gibson in that an affordance is what an artifact ``offers the animal, what it provides or furnishes'' \cite[p. 127]{gibson1979ecological}. AI models are (very!) complicated artifacts, and many of their affordances are initially ``hidden'' (not visually perceptible); however, this does not preclude their characterization nor mean they are perfectly designed for in current systems.} What these visions of interaction share is the promise of AI models to \emph{ground} communication between humans and machines \emph{in the human's preferred way}, rather than the machine's and its designers'. 

We call this affordance of human-computer interaction \emph{dynamic grounding}. The adjective \emph{dynamic} calls out the contrast with the largely \emph{static}, machine-over-user power differential of establishing common ground~\cite{clark1996using} instantiated in past software~\cite{li2023beyond}. In \emph{dynamic grounding,} the user grounds communication with the AI in whatever way is at-the-moment relevant to them. This ‘way’ may be ad-hoc notations, interactions, bespoke interfaces, etc. It is often \emph{ephemeral} or \emph{disposable}, such as assigning meanings to shapes in a drawing which no longer hold outside that particular conversation. %
As an example, imagine a user describes an algorithm to an AI by writing it out in Haskell pseudo-code. The next day, they communicate how to improve the program via a sketched tree diagram of a recursive data structure. The AI communicates the changes via an dynamically generated interactive widget of the tree diagram, which the user edits to modify, rather than the code itself.  \emph{What} the human and AI are communicating about---their \emph{joint project}---remains the same, but \emph{how} they communicate this information---their notation or lingo---changes on a whim. Beyond emphemerality, it could also be that a user establishes a notation or interaction---really, assigns a semantics to a `syntax'---that holds for the rest of a joint project (e.g., in mathematical research symbols are defined in early meetings then referenced in later meetings without explicit definition). Thus, the primary characteristic of dynamic grounding is the reversal of traditional power dynamic between user and machine when establishing common ground. Through AI-enabled interfaces, the user can take the lead in establishing common ground and define the rules through which that ground is contested.\footnote{By a shift of power we do not mean the user's power is absolute (nor may this be a valuable goal, as friction with constraints can prove generative \cite{kaptelinin2006acting}).} %

\subsection{Constructive Negotiation} \label{constructive-negotiation}

\begin{figure*}[t]
  \centering
  \begin{subfigure}[b]{0.7\textwidth}
    \includegraphics[width=\textwidth]{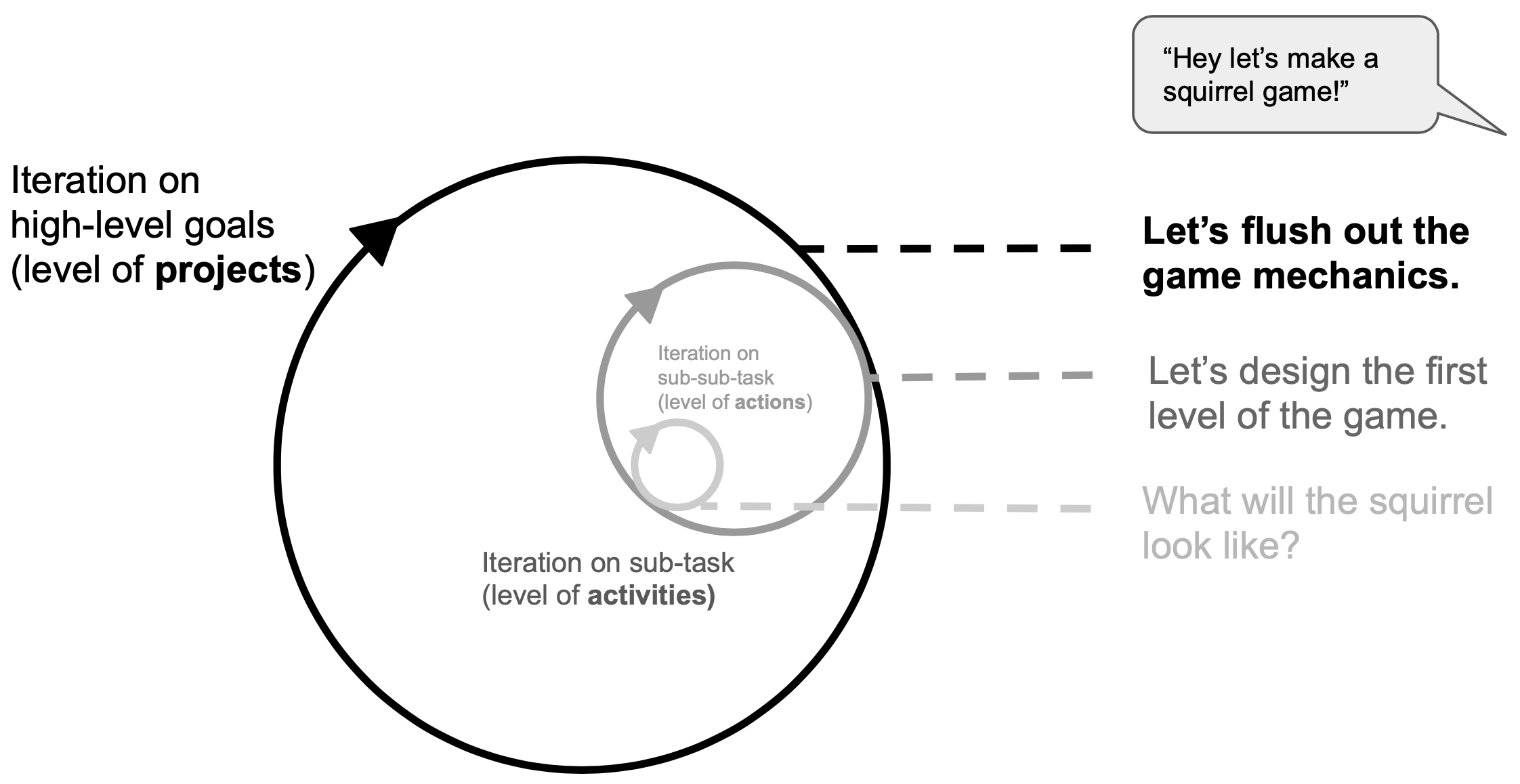}
    \caption{Fractal Design Spiral}
    \label{fig:sub1}
  \end{subfigure}
  \hfill
  \begin{subfigure}[b]{0.24\textwidth}
    \includegraphics[width=\textwidth]{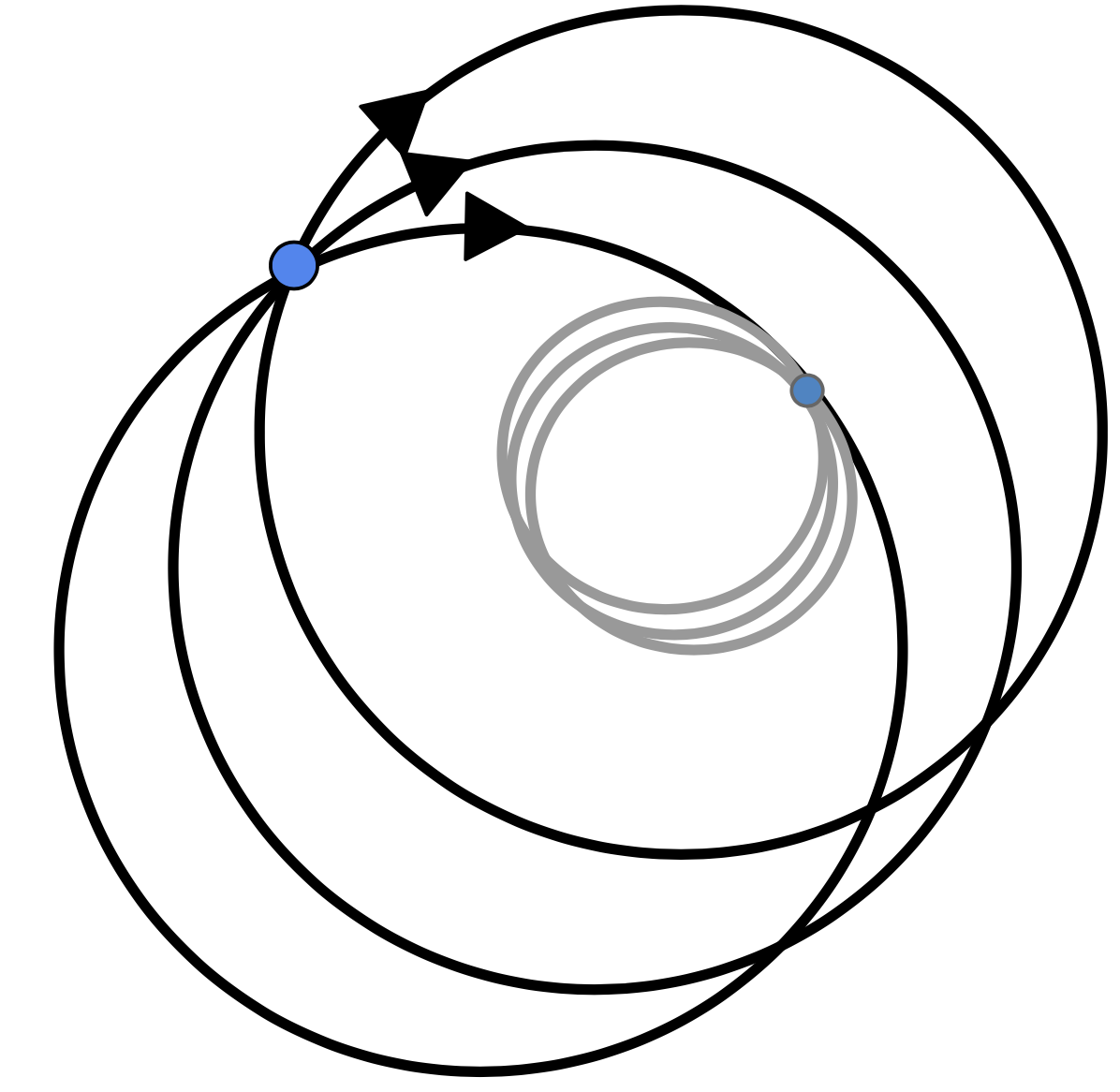}
    \caption{Design decision points}
    \label{fig:sub2}
  \end{subfigure}

  \caption{(a) The fractal design spiral (FDS). Design iteration moves from a high-level, abstract discussion of a \emph{project} and its \emph{goals}, to lower-level \emph{activities} and finally \emph{actions}. The AI needs to keep track of `where' one is in the spiral and integrate outcomes made at lower levels back into upper levels (for instance, the AI should `remember' choosing a sprite for the squirrel protagonist while prototyping the first level, and `integrate' it at the project-level as well). The \emph{motives} of users may emerge throughout this fractal design process (according to activity theory motives are rarely prespecified or even well-understood by users themselves \cite{kaptelinin2006acting}). (b) Disambiguation with an AI, mapped onto the fractal design spiral. The user is at the blue decision point, choosing between options. Deciding on project-level concepts and goals ``covers more distance'' through the design space, while nested sub-tasks cover less and less ground (for instance, choosing to center a game around a squirrel's life or a shark's, versus where to place a collectible in level four). Thus, AI `antagonism' and constructive negotiation is arguably more important at higher levels of abstraction (related to Fig.~\ref{fig:conflict-vs-level}).}
  \label{fig:spiral}
\end{figure*}

Helping users communicate their intent to machines while minimizing the effort or `translation work' involved \cite{arawjo2020write} has been a core concern of HCI design for decades. The field of user experience design even invented an adage {\it ``the user is always right''} to express how computer interfaces need to serve users and their intent. Consequently, the most popular AI models are framed as ``assistants'' and trained to be subservient and sycophantic \cite{sharma2023towards}. Attention is being paid to using AI models to help infer and disambiguate user intent, either in one-shot or turn-based chat interactions \cite{ma2023beyond}.

Intent elicitation is powerful and by no means easy to design for. Yet there are several problems with this vision of human-machine communication. First, the notion that a user has a coherent ``intent'' or ``plan'' that they merely need to communicate to the machine is rarely the case. Early work in HCI showed that humans do not follow prescribed plans but rather take actions in response to evolving contingencies~\cite{suchman1987plans, kaptelinin2006acting}. %
A related problem is that the user is \emph{not} always right. In fact, as designers, we are more likely to be at first wrong, or at least subpar, and need to rapidly iterate until we reach a suitable solution. The entirety of research through design (RtD), ``a process of making and critiquing artifacts,'' rests upon this premise \cite[p. 167]{zimmerman2014research}. 
Conflict and compromise are therefore central to good design. 

An array of literature across fields attests to the benefits of conflict, such as organizational management, software engineering, and intercultural communication~\cite{johnson2000constructive, jehn1995multimethod, janis1972victims, gobeli1998managing}. Each field shows that moderate amounts of conflict can be beneficial, provided that they are managed responsibly~\cite{gobeli1998managing}. Designers have long known this---art schools hold design critiques; game developers playtest their work with newcomers; fiction writers get ``beta readers'' to read their work before print; UX researchers conduct need-finding interviews and build lo-fidelity prototypes to identify issues early before significant investment has been made in implementation. These ``design critiques are not only about aesthetics, but also about concept, systems, meaning, and culture''~\cite{Muller2022}. Critique does not just strengthen ideas, but averts disasters caused by thoughtless agreement and complacency~\cite{janis1972victims}.

During collaborations over joint projects, then, \emph{negotiation} is \emph{constructive}, even necessary. Instead of framing human-AI interaction as a one-way street, the machine should \emph{push back}, \emph{constructively negotiate} with the human to consider design aspects that they had not yet anticipated, whether in a design's functionality, form, or anticipated reception.  %
Merely stating that constructive negotiation is beneficial, though, is not enough. As mentioned above, conflict is only generative when it is in \textit{moderate amounts} and \textit{managed responsibly}. Too much or too intense the conflict, outcomes suffer; no conflict and problems are not identified or team members' unique information is suppressed \cite{cao2021my}. The benefits of conflict also differ depending on the level of activity \cite{jehn1995multimethod}. Let us unpack these findings and their implications. 

Jehn separates activities into two types: \emph{non-routine} and \emph{routine} \cite{jehn1995multimethod}.\footnote{These roughly correspond to activity theory's level of activities, which go from ``activities'' to ``actions'' to ``operations'' \cite{kaptelinin2006acting}. Sometimes, operations are so rote that domain experts may not even be aware they are performing them.} %
Whether conflict is beneficial depends on the type of activity and level of abstraction. To visualize this, see Figure~\ref{fig:spiral}, which represents design processes as a fractal design spiral. The size of the circle represents the level of abstraction (level of activity), with bigger circles representing higher levels of abstraction. Jehn found that when groups were performing ``non-routine'' tasks---tasks that ``require problem-solving, have few set procedures, and have a high degree of uncertainty''---conflicts can be beneficial and even improve outcomes \cite{jehn1995multimethod}. In Fig~\ref{fig:spiral}, these tasks generally occur in the higher levels of activity. So too do we see that, at higher levels of activity, each decision point significantly changes the project direction and scope (schematically they also modify the design \emph{space} significantly). 

\begin{figure}
    \centering
    \includegraphics[width=0.7\linewidth]{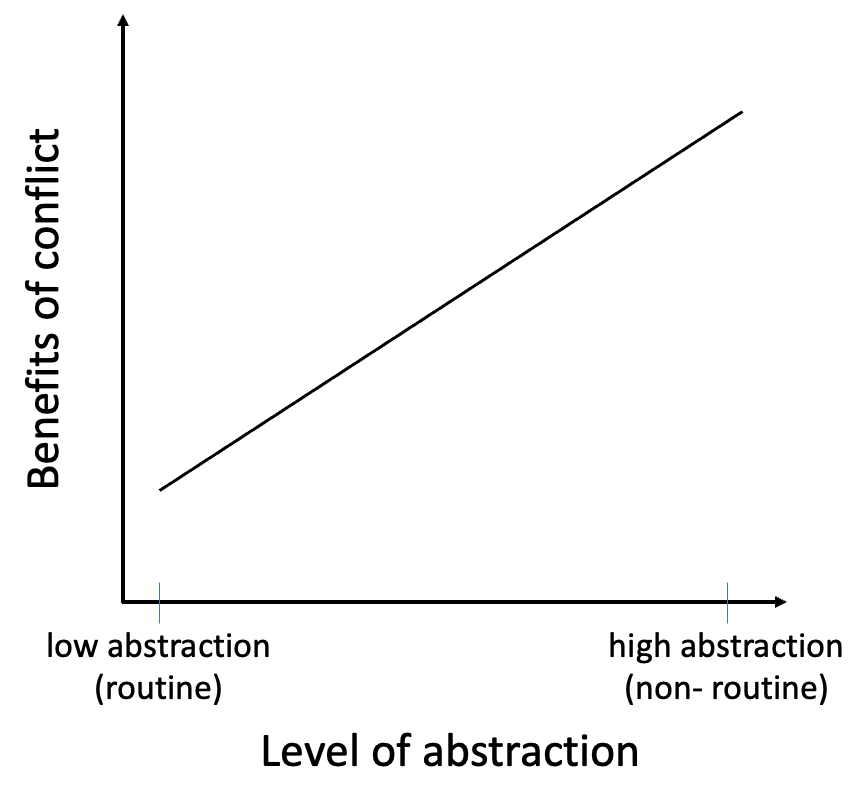}
    \caption{Benefits of Conflict vs. Level of Abstraction. Design tasks and decisions at lower levels of abstraction are more routine tasks that do little sway the overall concept (e.g., high-fidelity choices like UI button color). Activities and decisions at a high level of abstraction are less routine, demanding more creative energy and benefiting more from conflict (e.g., brainstorming).}
    \label{fig:conflict-vs-level}
\end{figure}

By contrast, Jehn found that ``routine,'' repetitive tasks that are generally ``done the same way each time'' do not benefit from conflict. In Fig~\ref{fig:spiral}, routine tasks generally occur in the inner spirals. At lower levels of activity, a local decision point doesn't sway the project significantly in the design space. For example, defining a function to reverse a string is routine, and would not benefit from conflict, while brainstorming a game design about a squirrel's daily life is non-routine, and can be beneficial. Thus, in early-stage and more formative iterations of design, we suggest an AI should be more ``antagonistic''~\cite{antagonistic_ai}; in lower levels of abstraction, later in a project, and during more routine operations, the AI should be less antagonistic.

How the AI manages conflict is also important. Drawing from~\citet{gobeli1998managing}, we suggest the AI should facilitate confrontational and "give and take" strategies of conflict resolution --- in other words, "collaborative problem-solving to reach a solution," or otherwise "reach a compromise solution which both parties can accept." Strategies of avoiding the issue ("withdrawal"), minimizing the differences ("smoothing"), or forcing the adoption of one side's solution ("forcing") should be avoided~\cite{gobeli1998managing}. Said differently, an AI that refuses to accept the user's ideas--strictly being disagreeable---is not helpful. Rather, the AI should be able to provide healthy conflict, and negotiation over disagreements should proceed with the ultimate goal of synthesizing perspectives.

Unfortunately, although the affordance of constructive negotiation is possible using today's AI models, it is suppressed. Training processes like reinforcement learning have resulted in sycophancy and hinder chat models' ability to establish common ground with users through interaction~\cite{shaikh2023grounding, sharma2023towards}. Shaikh et al.~show that popular LLMs are presumptive grounders, ``biased towards assuming common ground without using grounding acts''~\cite{shaikh2023grounding}. Future advances in training processes or alternative training data could change this. The sycophancy present in models also intersects with concerns of fairness, where an objective like Anthropic's ``harmlessness'' can easily be conflated with deference~\cite{antagonistic_ai}. Directives of harmlessness lead to AI chat models that are highly averse to criticizing users' thoughts and ideas, to the point of agreeing with them on factually incorrect information~\cite{sharma2023towards}. All this acknowledged, recent work on priming AI chatbots to help humans practice difficult conversations shows the potential of `antagonistic' AI~\cite{shaikh2023rehearsal, antagonistic_ai}.

\subsection{Sustainable Motivation}

Many projects worth doing take time, requiring persistence and struggle. It is important to manage and maintain motivation across long time frames~\cite{brooks1995mythical, beecham2008motivation, melo2012developers}. Successful management requires planning and contextual understanding at every level of project activity. Project management frameworks, like Agile in software engineering, require detailed documentation and statistical measurements to constantly modify and adapt to changing project environments---e.g., changing timelines,  requirements, and technical constraints. Because of how they embed social and cultural knowledge, AI tools can help understand user context, improvise current task(s) and plan(s) through the integration of new contextual information, and provide high-level project management as the project evolves. 

Prior systems did not share enough \emph{common ground} with users' social, cultural, and domain-specific concerns. For instance, for the expression \emph{``my daughter is sick and I need to work from home today to take care of her''}, prior NLP models could not understand the social context and intelligently apply them to adjust user plans. However, LLMs \emph{can} ``understand'' (or at least appear to understand) social context. They can therefore support the user in ad-hoc planning, improvisation, and motivation over the course of a project, provided this affordance is designed for.

We define \emph{sustainable motivation} as the affordance of AI systems to provide long-term support for users to successfully complete their projects and achieve their goals through ad-hoc improvisation and reaction to real-world contingencies. For instance, as a project starts, the AI can decompose a high-level abstract plan into a detailed timeline based on the project deadline. %
Whenever a user picks up a new task or a previous task after an interruption, the AI can help the user quickly understand where they left off, what has changed since then, and a plan on how they should proceed. If the user is not interested in programming for the day, or the user is traveling in a plane without access to the internet, they can communicate these constraints to the AI, who can adjust plans (e.g., tasks that don't require an internet connection where they sketch using a digital pen) and even prepare the system for future events (e.g., pre-downloading a Figma Jam file from the internet).

Beyond responding to external factors, the AI could also react to feelings when explicitly conveyed by the user. If the user communicates that they feel down or mentally drained but still wish to contribute to a project, the AI may narrow the type of activities suggested and adjust the tone of its response (i.e., the degree of antagonism may go down). For instance, for a software developer who feels overwhelmed, the AI could suggest the user solve a GitHub Issue that it judges as likely solvable within an hour, akin to self-sourcing ease-in microtasks~\cite{cai2016chain}. Relatedly, the AI could augment contextual information to help users to successfully \emph{return-to-focus} ongoing tasks~\cite{kersten2006using}. This information includes where the user left off during the previous session, what changed between the current and the previous work session, and other task-relevant information and artifacts. Finally, the AI could recommend transitions and breaks at opportune moments to optimize user happiness and productivity~\cite{kaur2020optimizing, hales2023juggling}, and adjust based on information about the user's broader context, such as Calendar, email, and text messages.

Prior research on \emph{task-switching} and maintaining \emph{flow state}, provides us with the kind of support that is needed to sustainably motivate users over the long development time frame. \emph{Flow}~\cite{csikszentmihalyi1975flowing} has been defined as an enjoyable experience engaging in a task that is appropriately \emph{challenging} and \emph{motivating}. Project planning and time management are critical to maintaining motivation and flow state over long durations~\cite{noda2023devex}. %

\section{The Story of Squirrel Game: A Vision of Designing with AI} \label{story}

Having established these three affordances, we then looked to imagine a future AI tool to support design processes, situated in a real-world example. We chose 2D game design as a context, as it requires the programming of complex systems with multiple media modalities, and is widely understood and broadly applicable. Adopting design fiction \cite{bleecker2022design, linehan2014alternate} and scenario-based design \cite{nathan2008envisioning}, we imagined an interaction through an aspirational fictional story (writing, illustrations, sketches, etc.) centered around our imagined, ideal design tool.\footnote{Blending HCI design with science fiction, design fictions are a ``world building'' activity which utilizes ``diegetic prototypes'' (fictional ones), the purpose of which is to open up a space for discussion~\cite{coulton2017design, lindley2015back, bleecker2022design}. A design fiction is ``(1) something that creates a story world, (2) has something being prototyped within that story world, (3) does so in order to create a discursive space''~\cite[p. 210]{lindley2015back}.} This story serves as a point for further discussion and object around which to mobilize. We ideated workflows separately at first, then came together to share thoughts and negotiate ideas. Over the next few weeks, we met in person on a whiteboard or over scratch paper, often for hours at a time. Throughout our design sessions, we reflected on how our imaginations may be conditioned by past practices and limitations. We were also wary of being too specific, wanting to evoke rather than prescribe.

While ideating, we kept in mind current developments in AI and technical limitations, following design fiction in ``work[ing] in the space between the arrogance of science fact, and the seriously playful imaginary of science fiction''~\cite[p. 8]{bleecker2022design}. Our goal was to ground our design fiction by real LLM outputs, such that, with ample time and resources, we knew it was possible to actually implement. We achieved this by prototyping LLM's ability to negotiate and plan via both direct chat interactions and prototyping prompt chains using ChainForge~\cite{arawjo2023chainforge}. We incorporated real outputs into our story. For instance, the suggestion below for a fox villain came from a suggestion of a GPT-4 model primed to be antagonistic---harshly, but constructively critical to the user's design ideas. %
We also tested the ability of models to integrate suggestions at lower levels of abstraction (e.g., while working on a single level) into a high-level design plan.

\newcommand{\think}[1]{\emph{#1}}
\newcommand{\quo}[1]{``{#1}''}
\newcommand{\ai}[1]{``\texttt{\small{#1}}''}
\newcommand{\grounding}[1]{\textcircled{\tiny{G}}}
\newcommand{\negotiation}[1]{\textcircled{\tiny{N}}}
\newcommand{\motivation}[1]{\textcircled{\tiny{M}}}

In the story below, we prepend \grounding{} to indicate instances of dynamic grounding, \negotiation{} for constructive negotiation, and \motivation{} for sustainable motivation. These may not be the \emph{only} instances---\motivation{}, for instance, is more holistic. In Section~\ref{story-debrief}, we will reflect on this story and suggest implications for technical implementation.

\subsection{Scenario}

\begin{quote}
    \emph{Game development is often a laborious process---even for hobby developers. It requires learning a slew of game design tools, detailed planning, etc. In the near-future of designing with AI support, we show how AI streamlines the process of game design and development through the eyes of Alice, a 12-year-old aspiring game designer.}
\end{quote}

\setlength{\parskip}{0.4em} %

\begin{figure}
    \centering
    \includegraphics[width=0.6\linewidth]{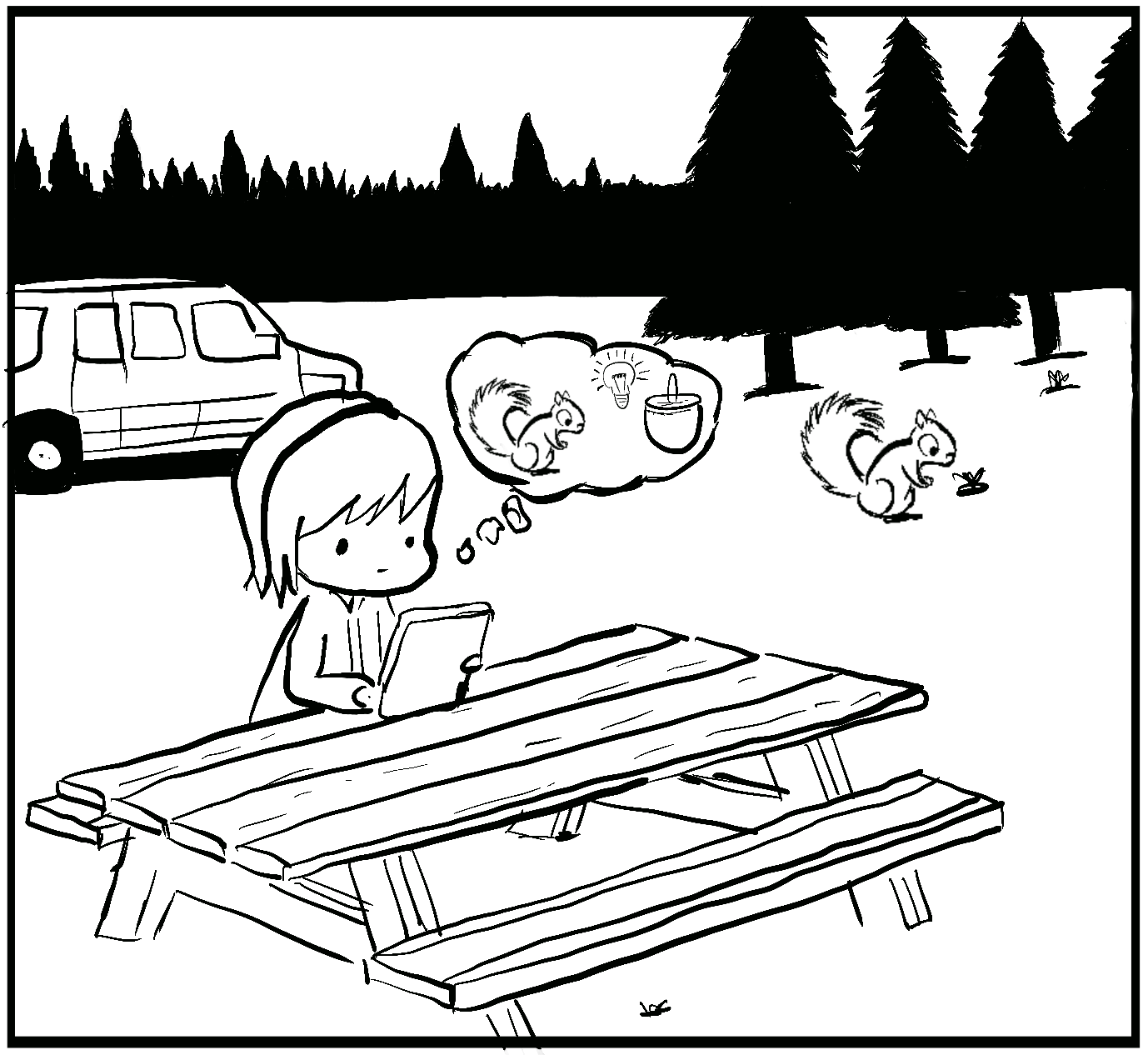}
    \caption{Alice notices a squirrel while sitting on a picnic bench. An idea flashes her mind: \quo{I want to make a squirrel game.}}
    \label{fig:comic-intro}
\end{figure}

Alice is traveling through Yellowstone National Park with her parents and younger brother. Having taken a life science class this past spring, she's fascinated by the diversity of wildlife---bisons, elks, eagles. She wants to wander off and explore on her own, but her parents say it's too dangerous. \think{Figures.}

They drive into a camping rest stop, and her parents go with her brother to get lunch. She waits at a picnic bench by the car, taking in the scenery. A glimmer catches her eye. A flying, furry---something? It glides across the road into a tree and scurries down. \think{A squirrel?} The rodent runs right by her, then burrows into the ground to emerge with nuts.

An idea flashes in her mind. A way to capture this scene and to convey its complexity, she gets out her iSwitch 64 tablet and opens Game Jammer, a tool her friends in the game development club showed her last month. 
She starts the chat conversation with Jarvis, the app's AI agent, by saying: \quo{I want to make a squirrel game.}

\ai{A game about squirrels? Or a game for squirrels?} Jarvis asks a clarifying question. The AI has a habit of doing this, Alice has noticed, whenever she's starting a new project. 

\quo{About squirrels, silly. I want to make a game about squirrels collecting nuts for the winter. And all the roads and obstacles they encounter. And maybe how fun it is to be a squirrel, too.}

\ai{OK, let's explore that. What's the player's goal in this game? Say, the first level.}

Alice thinks a moment. \quo{They need to collect nuts.} %

\negotiation{} \ai{I can create that game, but it sounds a bit bland.} Alice rolls her eyes. \ai{It needs more depth and challenge. Let's make the idea more original.} Jarvis lists three suggestions, including adding power-ups, introducing a villain character, and even a starving squirrel family that the main character needs to feed.\footnote{These suggestions may span a range from typical game mechanics to very unusual. All suggestions here were actual LLM-generated suggestions the authors saw while prototyping with an AI primed to be critical of user ideas. Section~\ref{sec:tropes} speaks about the value and concerns of AI suggestions.}

Alice blinks. She didn't even think of a story to the game. \quo{Oh, that's a great idea---a villain! What if we make it a fox? Don't foxes hunt squirrels?}

\ai{They do indeed, according to Wikipedia. Perhaps the fox chases the squirrel if it is too slow?}

\negotiation{} \quo{Sure! Let's go with that.} Alice is excited. She wants to add power-ups too, and they have a negotiation about how to add a power-up that can remove cars from the road. Alice thinks about allowing the squirrel to blow up cars, but Jarvis considers that a little violent. They compromise on a lightning bolt that strikes a tree, felling it across the road, preventing cars from passing. 

\quo{Enough ideas for now! Let's make it!} Alice's family is still gone, but she wants something done before her brother comes back and disrupts her concentration. 

\ai{Ok, here's what I've understood.} Jarvis outputs a list of features the game will have, from collecting nuts to the fox villain. Alice reviews the list briefly. \quo{Hmm, ok. What should we do now?}

\ai{Here's some steps.} Jarvis outputs the start of a plan, including prototyping the first three levels and choosing an art style. \ai{Let's make level one. Should I generate a level based on what we discussed?}

\begin{figure*}[t]
    \centering
    \includegraphics[width=1\textwidth]{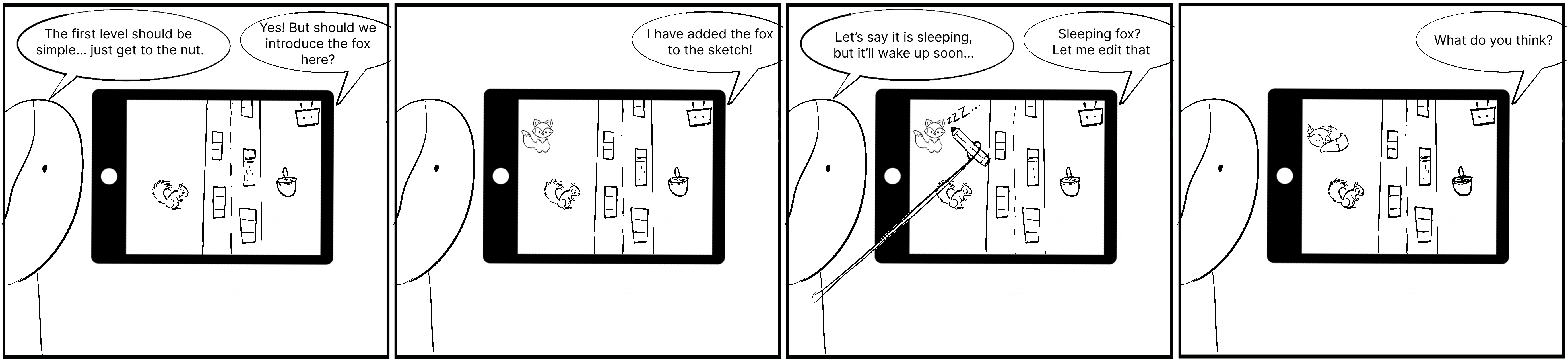}
    \caption{Alice draws the first level of Squirrel Game. The AI reminds Alice of a fox character decided in an earlier design conversation. The graphical style of the AI's suggested fox is grounded in Alice's sketchy style. When Alice adds ZZZ's atop the fox to represent sleeping, the AI edits the fox sketch to be sleeping, again mimicking Alice's style.}
    \label{fig:comic-dg}
\end{figure*}

\grounding{} Alice slips her stylus from the iSwitch's holster. \quo{I'll draw it! Give me a canvas.} Jarvis switches to a fullscreen canvas. Alice sketches a squirrel, an acorn beneath a tree, and the road between them. \quo{The first level should be simple, right? Get to the nut, and do not get squashed by traffic.}

\ai{You've drawn a squirrel, a road, and a nut beneath a tree across the road. Is that correct?} While Jarvis talks, soft colors highlight the elements of the sketch he refers to.

\quo{Yup. What do you think?}

\grounding{} \motivation{} A fox sketch appears on screen in Alice's rough style, right behind the squirrel. The fox is gray and not fully black, expressing a suggestion from Jarvis. Jarvis asks: \ai{What about the fox villain? Should we introduce it here?}

\grounding{} \quo{Oh, right! Let's say it's sleeping but it'll wake up soon.} Alice drags the fox up with her finger and draws ZZZ's coming from the fox. Jarvis modifies the fox sketch to be curled up with eyes closed. \ai{How soon? How many seconds?} \quo{Ten!}

\ai{Ok, want to play it? Or should we work on the idea more?}

Alice looks up. Still no sign of her family, but they're coming back soon. She can't work in the car, she'll get motion sickness. And she does want to hike today. \quo{Make it! Quickly. I have to go soon.}

\motivation{} Jarvis ``thinks'', spending time creating a playable prototype of the squirrel game. It considers asking Alice what control scheme and art style she wants---however, Alice is in a rush and wants results quickly. Eliciting more intent would just slow things down. Looking at the graphics and controls of the other games Alice likes to play on her iSwitch 64, it decides on a poppy, colorful aesthetic and a digital thumbstick control, and pins the questions for later. %

The screen flickers, and a title card with ``Squirrel Game'' pops up. A button ``Play Start'' is below. Alice touches it. 

Alice plays the game, using the thumbstick to move the squirrel. She drew a short road, and a single car goes across it every few seconds. She crosses the road easily and grabs the nut. ``Nut-tastic! Level one complete'' appears on the screen. \quo{Hmm, it's way too easy. Maybe the car goes faster or there are more cars? What if the cars go crazy fast?}

\motivation{} Jarvis considers starting a negotiation. However, once again, it remembers that Alice is in a rush. \ai{I can add that}, says Jarvis. The level resets, with more cars zooming down the road. Now it is not so simple to pass. 

Alice plays it, dodging the cars. \quo{Better! Can we work on level two?}

\quo{Alice, what are you doing? On the iSwitch again?} Alice's mother appears beside her, shaking her head. \quo{It's time to go see Old Faithful.} Alice puts away her iSwitch and joins her family in the car.

While Alice resumes her trip across the national park, Jarvis is working hard in the background to generate the designs for further levels of the game. 
The next morning, Alice opens her iSwitch 64 tablet with a warm mug of tea in her hand. She taps Game Jammer. \quo{Hey Jarvis!}

\ai{Welcome back Alice! How's your morning going?}

\quo{Oh my god you wouldn't believe it---I saw a fox on a hike yesterday! Can't wait to resume work on our game.}

\motivation{} \ai{Of course! There are some things for you to review. In our session yesterday, you designed and played the first level.} A sketch of the first level appears on screen. \ai{I wanted to ask you about the art style for the game, but you were in a rush. Is the pop-art style okay, or would you like a different style?}

\quo{Hmm, let's decide on the style later. What you had was good.}

\motivation{} \ai{Okay, I'll mark that for later. While you were gone, I generated some designs for levels 2 and 3. Do you want to try play testing them?}

\quo{Oh yes! Let's do it!!}

Alice tries out the next level of the game. At this level, there are bigger vehicles like trucks that make the timing of crossing the road tricky---she likes that challenge. After the third level, though, she begins to think that the game is getting monotonous. \quo{Jarvis, though more vehicles make it more tricky, the game feels the same pretty fast. Just adding more lanes and trucks won't cut it.}

\motivation{} \ai{Maybe this is the right time to think about introducing the lightning power-up we discussed?}

\motivation{} \quo{Oh! I completely forgot about that. That will be perfect.  Give me a canvas.}

\grounding{} \ai{Here is a sketch of the third level's design.} Jarvis presents a sketch in a similar style to how Alice sketched level one, rather than the high-fidelity pop-art aesthetic at runtime.

Alice draws two thunderbolts on the road on top of the existing design for the third level. One is in the center of traffic, but another is close to the squirrel's side of the road. \quo{What do you think?} 

\grounding{} \negotiation{} \ai{It looks easy to get that one.} Jarvis indicates the thunderbolt closest to the road with a soft circle overlay. \ai{Want me to remove it? Or should we discuss changing the design? There are alternative ways to obtain and activate power-ups in games that we can discuss.}

\quo{Hmm I'll remove it.} Alice erases it. \quo{That's better. Let's try like this!}

\ai{Just a moment, I'll get the level ready for you.}

\grounding{} Based on Alice's annotations on the level map, Jarvis modifies level 3 and starts the level for Alice. Alice dodges through traffic, narrowly missing a truck to collect the power-up. A lightning bolt strikes a tree near the top of the screen, and it falls across the road. To Alice's surprise, a ``road closed'' sign also appears at the opposite end of the road.\footnote{In any AI system for design there is a trade-off between the AI making assumptions and resolving problems itself, and explicit discussion with the user to disambiguate intent and resolve emergent problems. Here, Jarvis encountered an emergent logical problem with implementing a part of the design plan, and resolved it without user input, as the user was away.} The traffic disappears. Alice is now free to go collect the remaining nuts. However, she realizes that stopping the traffic has made collecting the nuts very easy. 

\quo{I like the power-up, but what if it makes the game too easy when they use it?}

\ai{That is a classic problem when designing a risk-reward system in games. The risk and reward should be well balanced. I have a few ideas, if you'd like to hear them.}

\quo{Tell me!}

\ai{The lightning bolt was not as easy to implement as I had thought. It required a road sign to block the opposing traffic, otherwise, the vehicles would bunch up. It is also fantastical and removes a central challenge of the game, which is the traffic. I think there is more we could do with the squirrel alone that tests player skill. What do you think?}

\quo{I like the power-up though! But what if we change it? What if the cars do bunch up? Could the squirrel slip through? Maybe it is still difficult since cars are still on the road, even if they are not moving. And maybe we can make the roads long and curvy in later levels, too.}

\ai{I can try that. I'll remove the road closed sign and bunch up the cars around the fallen tree.} 

\grounding{} Jarvis now generates a few level designs with curved roads and places the power-up strategically across the map. \ai{Here are some level designs to playtest.} Sketches of three potential level designs appear on the screen, each with a different curve and power-up placement. Alice taps one.

Alice loves navigating around complex curves on the roads. On the second attempt, she was chased by the fox when she took a lot of time to collect the power-up, and the bunched-up traffic from using the power-up blocks her in and prevents her from reaching the final nut. ``That sly fox!'' text appears, with a ``Try again?'' button. 

\quo{Dang, I totally blew it}, says Alice. \quo{I love the curved roads with power-ups though. Can you generate a few more levels like this? Maybe one has two roads to cross!} Alice continues to collaborate with Jarvis to settle on the final version of the level. 

After a few more days, Alice, back from her trip, visits her favorite game development club excited to show off her prototype. 

\quo{Hey Alice! How was your trip to Yellowstone?} asks John, a curly-haired boy with glasses.

\quo{Hi John! The trip was great. Funny, while I loved watching the geysers, I also spent some time making a new game---with Game Jammer. Would you like to try it?}

\quo{A nerdy holiday huh? This is what happens if you take your iSwitch with you on your holidays. I would love to try the game. Maybe later I can know more about your experience with Game Jammer---I'm very skeptical about it.}

Alice takes her iSwitch tablet out of her backpack and shares the playable prototype with John. After ten minutes, John has played all the available levels in the prototype. 

\quo{Pretty cool Alice! I have a couple of comments though. The controls are a bit slow to react especially when you change direction---it can be smoother. I also loved the lightning idea, but don't squirrels also climb trees? It's weird there's trees but no way to climb. Maybe they can be flying squirrels and that's another way to dodge traffic.}

Alice thinks back to sitting on the bench, seeing a squirrel glide across the road. Had she imagined it? She had almost forgotten about that. \quo{Squirrels can fly?}

John raises his eyebrows. \quo{Well, yeah! When I went to Yellowstone there are flying squirrels. Didn't you see any?}

Alice takes the iSwitch and instructs Jarvis---\quo{Hey Jarvis, what if we gave the squirrel flying powers?}

\ai{Do you mean flying like a superhero, or flying like a flying squirrel? A flying squirrel is a real species of squirrel that can glide between trees using a patagium, or flap between their arms. Here is a video of it.} A video appears with a squirrel jumping from a branch and gliding to another tree.

\quo{Woah!} says Alice. \quo{I think I saw that at Yellowstone! Let's go with the flying squirrel.} %

\negotiation{} \ai{That sounds like a unique twist. There are some things to think about. When should the squirrel be allowed to glide, and for how long? Is it from the start of the game, or unlocked at a later level? Does it use a stamina bar that builds up over time? Or can you do it only after climbing large trees?}

Alice looks to John and says with a grin, \quo{Sometimes he gets like this.} Alice thinks. \quo{Hmm Jarvis, I think it should be unlocked later. %
Maybe the squirrel can glide but only if they climb to the top of a tree. How difficult would this be to do?}

\ai{We would need to add a climbable tree and indicate that to the player. Perhaps there is a special tree graphic that indicates a tree is climbable. We could place it beside a road. Yet it could also reduce the challenge of the game if we're not careful.}

\quo{Hmm let's try it and change it if it's too easy. Maybe we introduce it in level four? And then we'll have to revise the levels. Can you modify the game to include it? We can figure out the details later. I would like to show it to my friend ASAP.}

\ai{Of course. I'll go ahead and make a new level four with the gliding mechanic. Would you like to confirm the plan, or make me do all the guesswork again?}

Alice smiles. \quo{Confirm.}

\ai{I will add a new tree type that is climbable. When the squirrel climbs atop it by running into it, they will glide in a player-chosen direction. This requires significant changes to the physics engine, and it will take a while to make all the changes. Sound good?}

\quo{Sounds good! Maybe we can try it tonight, and I'll show my friend at the next club meeting.}

\section{Discussion and Challenges to Realization} \label{story-debrief}

\setlength{\parskip}{0.0em} %

Our fictional story was iterated to feel as realistic as possible. We wanted Alice to feel real---not a concocted user in a lab study, but a child out in the world, full of eagerness and impatience. Although we describe a futuristic technology, we were wary of too much hand-waving and idealism in how Jarvis approaches situations, implements ideas, and resolves ambiguities in intent. It is plausible that something like Game Jammer could exist today if it were constrained to simple, 2D games. Even when constrained, however, how Jammer implements dynamic grounding, triggers negotiations and ensures they remain constructive, remembers past details, and sustains Alice's motivation is a massive endeavor for a team of full-time software designers and engineers to pull off. We believe such a system is within reach, so here, we want to sketch some thoughts on technical implementation and reflect on the story.

Throughout Jarvis's conversations with Alice, there is a layer that considers context before making a decision---whether to ask a clarifying question, how to phrase its response, or what response to give. This layer accomplishes sustainable motivation and is the most holistic of all the three affordances of AI model considered here. Both before and after design decisions are made, Jarvis also considers a negotiation---a critique of the design, push-back that it may (or may not) choose to present to Alice. Opportunities for negotiation intersect with turn-based intent elicitation, an affordance described in past work \cite{ma2023beyond}. Finally, Jarvis lets Alice take the lead in grounding communication and thereafter attempts to ground its communication in Alice's notation and lingo (e.g., sketching in her style, or even how it replies, avoiding complicated words and extraneous detail). Below, we elaborate on some system requirements (4.1-5) and wicked problems (4.6) for the implementation of Game Jammer. 

\subsection{Localization of position in the design space}

Understanding the level of abstraction of the task (or conversation) is vital for the AI to collaborate in the design process. 
We can view the levels of abstraction of the project as a fractal design spiral (Fig~\ref{fig:spiral}), where the bigger circles represent iteration over high-level abstractions and the smaller inner circles represent iteration on low-level abstractions. Design iteration moves from a high-level, abstract discussion of a project and its goals, to lower-level activities and finally actions. The AI needs to keep track of \emph{where the user is} in this conceptual space (i.e., iterating on one specific graphic versus amending high-level game mechanics) to successfully plan, negotiate, and execute tasks. For instance, when Alice is working on a lightning power-up, the AI must keep localizing the work to level three, but also identify the power-up as a higher-level abstraction that could apply across the game.

One potential representation of the project that can help AI understand the level of abstraction and behave accordingly is a task hierarchy. Each project has a hierarchical list of tasks, where the highest level of tasks represents the broad components of the project. In \emph{Squirrel Game}, the top-level tasks can be conceptualization and design, game engine, game assets, audio, playtest, etc. Each high-level task can have multiple lower-level subtasks and so on. This task list will continuously evolve where both the user and the AI will add, modify, or delete tasks based on their negotiation. 

\subsection{Intent elicitation and disambiguation}

User intent, a high-level goal, is often ambiguous, only partially observable, and evolves throughout the project. For the AI to successfully understand, guide, and execute toward the user's goal, the AI should be able to successfully elicit and disambiguate the user's intent. The process for intent elicitation, also dubbed as \emph{finding common ground}~\cite{clark1996using} or \emph{intent prediction}~\cite{qu2019user}, involves developing a concrete model of user intent by understanding both explicit and implicit context of the user using founding acts such as \emph{clarification}, \emph{acknowledgment}, and \emph{follow-up}~\cite{clark1996using}. In our story, when Alice shares her intent to create a squirrel game, Jarvis asks clarifying questions about the gameplay to establish a shared model of what the game is. Another example is when Jarvis brings up art style to elicit clarification from Alice on the second day. 

Intent disambiguation is becoming standard practice in systems that support human-AI collaboration \cite{ma2023beyond, mu2023clarifygpt}. CLARA~\cite{park2023clara} infers the clarity, ambiguity, and feasibility of the user command using an uncertainty or ambiguity estimation with LLMs. We can also borrow a wealth of research on disambiguation and elicitation by asking clarifying questions from the information retrieval and dialogue systems fields~\cite{dhole2020resolving, keyvan2022approach, krasheninnikov2022assistance, min2020ambigqa, rao2019answer, trienes2019identifying}. From localizing the user's ``position'' in the design space~(Fig.~\ref{fig:spiral}), the AI may also be more proactive at resolving ambiguity when at higher levels of abstraction, as high-level design decisions often have wide trickle-down effects.

\subsection{Designing for constructive negotiation}

To elevate the role of the AI to a collaborator, the AI agent needs to go beyond intent elicitation and be capable of pushing back on user ideas. Priming AI to be antagonistic may benefit users to self-reflect, strengthen ideas, and even escape the status quo of user beliefs~\cite{antagonistic_ai}. In design, conflict can nudge users to think beyond their intent and explore alternate paths to ensure they spot flaws early and quickly. Jarvis' comment that Alice's idea was \quo{a bit bland} provoked her into deepening her design concept. A second negotiation around Alice's bomb power-up idea asked her to consider whether a less violent approach was more compelling.

Practically, to make use of constructive negotiation, the AI must be capable of being antagonistic, and also explore alternate paths that are against the user's ideas or commands. But just antagonism alone is not productive---the AI has to know \emph{when} and \emph{in what context} to enact a negotiation, and \emph{how} to manage it.
In Section~\ref{constructive-negotiation}, we cited literature showing that for conflict to be constructive, it must be in \emph{moderate amounts}, \emph{managed responsibly}, and is \emph{dependent on the task}. Non-routine, creative tasks (at high levels of abstraction) benefit the most from conflict, routine tasks the least (``actions'' in Fig.~\ref{fig:spiral}). Figure~\ref{fig:spiral}b shows the decision points where conflict and negotiation will lead to picking one of the many possible directions (circles). When the level of abstraction is higher (bigger circle), choosing a decision point significantly changes the design space compared to a lower abstraction level, where the circle of influence is smaller.

In the story, we saw tensions between sustainable motivation and constructive negotiation---Jarvis had to \emph{decide} whether to begin a negotiation, or critique, of user input, and this decision was sometimes augmented by time constraints (urgency) conveyed by Alice. When decisions must be made urgently, the AI system decided to hold fewer elicitation rounds, made more assumptions about Alice's intent, and triggered fewer negotiations. We can imagine a threshold must be reached to trigger a negotiation that is a complex function of many factors beyond urgency alone (i.e., if a user's design idea enacts a harmful stereotype, even if the user is in a rush the AI could be justified in pushing back). In practice, responsible antagonistic AI also involves considerations of consent, context, and narrativization \cite{antagonistic_ai}. Recent systems like Rehearsal~\cite{shaikh2023rehearsal} provide an example of using clever prompting techniques with guardrails to generate conflicts that can help users to pursue alternate conversation paths using counter-factual \sayit{what if?} scenarios. 

\subsection{Planning}
Efficient management and planning of the design process by AI will considerably improve the productivity of the users. For this, the AI has to dynamically adapt to the state of the project, user context both external and internal to the project, etc. On the second day of our story, Jarvis had planned to discuss the art style of the game but decided to table it temporarily due to Alice's interest. The AI can not only plan and allocate tasks based on the development timeline, but also optimize for the user's in-the-moment interest, and also optimize for user motivation. 
To successfully implement this, we can borrow research from automated project management, and AI-assisted project management. Though current research focuses on data-driven approaches to optimize resource allocation, improve risk assessment, etc.~\cite{taboada2023artificial, prifti2022optimizing, auth2021conceptual, gil2021application}, Barcaui et al. suggest that LLM-assisted project management can be moderately effective in resource planning, quality planning, risk management, and more; the drawbacks of LLMs centered around its lack of disciplinary and organizational context compared to human managers~\cite{barcaui2023better}. In the future, the AI may optimize user motivation and productivity by recommending transitions and breaks at opportune moments~\cite{kaur2020optimizing, hales2023juggling}, and re-plan tasks based on just-in-time information about the user's broader context, such as Calendar, email, and text messages. Yet while more context can improve sustainable motivation, there is a trade-off with privacy at individual, social, and organizational levels. Local LLMs and techniques like privacy-preserving prompt tuning can help with these concerns \cite{li2023privacy}. 

\subsection{Integration}
Every conversation with project collaborators or completion of a task often has ripple effects across other tasks both at higher and lower levels of abstraction. After elicitation turns to disambiguate intent, negotiations with users, and user decisions, the AI must be capable of \emph{integrating} these choices into the project plan or artifacts. In Squirrel Game, once Alice and the AI decide to introduce a fox villain as a gameplay element during the negotiation, the AI has to modify the high-level design plan and also (potentially) make additions and changes at lower levels of abstraction, such as creating a fox character as a design asset, adding the character to any existing levels, generating sound consistent with a chosen sound design aesthetic, etc. In another instance when Alice concretizes the power-up design (both mechanically and graphically), Jarvis must remember this abstraction upon generating further level design concepts.

Integration is complex even for simple projects like \emph{Squirrel Game} which involves the AI to remember, retrieve, and reflect necessary contextual information of the project.  A major challenge of current AI systems is to manage memory and context. The memory contains all the data necessary for the project analogous to the human brain, and the context is the \emph{localized knowledge} required for the model to complete the current task---analogous to working memory for humans.  Current language models cannot do this by default. However, the approach of \citet{park2023generative} suggests how AI agents can implement \emph{long-term memory}, \emph{memory retrieval}, \emph{reflection}, and \emph{planning} mechanisms necessary for integration. We can also achieve long-term generative agent memory using vector databases, and retrieve appropriate context using methods like retrieval-augmented-generation. Tools like CodePlan~\cite{bairi2023codeplan} perform repository-level code edits using LLMs using a multi-step chain of edits.

\subsection{Tropes and stereotypes in design support tools} \label{sec:tropes}

During design, a designer must choose between alternatives at many decision points and levels of abstraction (Fig.~\ref{fig:spiral}b). A human designer is limited by their cognitive load, prone to forgetting important details or missing aspects of design until they are noticed. Even generic ideas, such as suggesting a narrative for the game, could prove beneficial as they can cause the user to think about dimensions of design that they had not yet considered. %
Suggestions open up previously undiscerned or forgotten dimensions of variation within the abstract design space, as predicted by Variation Theory~\cite{marton2014necessary}. Both variation theory and activity theory argue that contrasts (what activity theory calls ``contradictions'') between new elements and existing designs can improve designs and support human development  \cite{kaptelinin2006acting, marton2014necessary}. %

However, the trade-off of suggestions from an AI is that they may be tropes that bias the designer towards the familiar. Game design in particular is a tension between remixing past tropes (of mechanics, aesthetics, narrative, etc.) with new and unexpected ideas. Tropes are design shortcuts---they can speed things along, motivate designers, and help onboard users---however, they can also lead to generic experiences.%
\footnote{For example, the platformer game \emph{Celeste} (2018) was not very new in its mechanics, but rather innovated with its narrative and music.}
In the story, we see this tension in the initial negotiation between Jarvis and Alice. Jarvis calls Alice's idea \ai{a bit bland} and suggests three ways to augment it: power-ups, a villain character, or a narrative motivation for the squirrel to collect nuts (for a starving family). All three were actual suggestions by GPT-4 in our prompt prototyping. While the last was the most unexpected to us, the first two are arguably video game tropes.  
These tropes can anchor or bias her, foreclosing alternatives (e.g., making the overarching narrative about ecological disaster and deforestation---perhaps the levels get harder as the squirrel's ecosystem is urbanized). Although priming the AI to be antagonistic seemed to lessen the generic-ness of suggestions, what the AI recommends during negotiation can still be a trope. This is a wicked problem~\cite{buchanan1992wicked}---there is no ``best'' solution to create the ``perfect'' design tool that is ``unbiased''---but the potential for the tool to bias users, anchoring them on mediocre but familiar design choices, is something to always consider. Despite all this, even tropes and other generic suggestions offer value: they contain contrasts to Alice's externalized prototype and/or current intent that helps her discern and consider aspects of the design she had not previously~\cite{marton2014necessary}). Designers of AI tools can consider this in their design choices: for instance, in how many alternatives the AI presents to users, or in where and how the interface asks for user input (e.g.,~\cite{ma2023beyond}). %

\subsection{Summary of system requirements}
In summary, here are key system requirements to implement Jarvis, the AI agent. Jarvis needs to keep in memory: 

\begin{itemize}
    \item a high-level plan for the game's design (mechanics, art style, story, feel, sound, etc)
    \item a model of who Alice is (her goals, preferences, desires, potentially background and age)
    \item a model of the current, external context (e.g., clock) \item any constraints on the game's development (e.g., target audience or platform, deadlines)
    \item a past history of interactions (conversations, actions Alice has taken, etc., a.k.a. a \emph{memory stream} \cite{park2023generative})
 \end{itemize}

\noindent Jarvis also needs a model of when and how to:
 \begin{itemize}
     \item ask questions to disambiguate intent
     \item start a negotiation or conflict, and to resolve it
     \item refer to past events or contextual information
 \end{itemize}

\noindent During the creation of the design, Jarvis needs to: 
 \begin{itemize}
     \item integrate decisions made at lower levels of abstraction (e.g., level one) into higher levels (i.e., the high-level plan)
     \item propose short-term tasks and a long-term development plan to successfully execute the project
     \item localize its ``position'' in the design space (i.e., what level of abstraction is the user currently iterating on?) and navigate ``through'' the design space (i.e., from level one to tweaking overall mechanics)
     \item ground its contributions and suggestions in the notation preference of the user (e.g., drawing a fox in a sketchy style akin to Alice's, or using similar language and terms to Alice)
 \end{itemize}

\section{Conclusion and Next Steps}

In this paper, we described three affordances of LLMs for supporting design work. Unlike prior work focusing on broad principles or top-down directives for human-AI collaboration (e.g., \cite{wang2020human,shneiderman2022human}), we 
took a bottom-up approach, centering a scenario of use through design fiction. Though our paper presented our ideas in a linear fashion, in reality, the affordances, story, and technical details were mutually constituted during our design discussions---the narrative fed into the affordances, the affordances into the narrative, prototypes of technical feasibility into the affordances, etc.

In the weeks since defining our three affordances, they have arisen again and again in our conversations with other human-AI interaction researchers. Often other researchers are trying to explain, without a term, the quality one of these terms describes. Unlike other papers we are a part of, then, our motivation was to provide clarity to future design discussions---others will come to these phenomena, but most likely in a disjointed fashion, in separate papers each focusing on one aspect or other. We hope this nomenclature can aid future communication over research---if we are aware of these affordances, we can then explicitly talk about using and exploring them in research, fruitfully bounding our conversations on where precisely large AI models can provide value compared to prior technologies.

It is important to remark that we do not consider it easy to implement or take advantage of each of our three affordances with current AI models. Significant scaffolding, technical innovation, and evaluation of interface designs must also be achieved to realize our near-future vision. Our method of using design fiction also had several limitations. While we erred against proposing a specific interface and implementation to evoke rather than prescribe, the process of designing an interface could have exposed us to questions and opportunities that we have not foreseen here. We also did not consider human-AI collaboration with \emph{multiple} users and stakeholders. While we believe the three affordances we mentioned will still be applicable, having multiple people may bring more challenges that need to be addressed.

\bibliographystyle{ACM-Reference-Format}
\bibliography{references}

\end{document}